# Dual Use Concerns of Generative AI and Large Language Models

**Alexei Grinbaum**
alexei.grinbaum@cea.fr
*CEA-Saclay/Larsim*
Gif-sur-Yvette 91191

**Laurynas Adomaitis**
laurynas.adomaitis@cea.fr
*CEA-Saclay/Larsim*
Gif-sur-Yvette 91191

*Abstract*
We suggest the implementation of the Dual Use Research of Concern (DURC) framework, originally designed for life sciences, to the domain of generative AI, with a specific focus on Large Language Models (LLMs). With its demonstrated advantages and drawbacks in biological research, we believe the DURC criteria can be effectively redefined for LLMs, potentially contributing to improved AI governance. Acknowledging the balance that must be struck when employing the DURC framework, we highlight its crucial political role in enhancing societal awareness of the impact of generative AI. As a final point, we offer a series of specific recommendations for applying the DURC approach to LLM research.

**Conflict of interest**
No conflict of interest to report.
**Funding**
This research was supported through projects *TechEthos* (grant number 101006249) and *MultiRATE* (grant number 101073929) funded by the European Commission Horizon program.
**Ethics approval**
No human subjects were involved in the study.
**Consent**
No data needing consent has been used.
**Data availability statement**
In this article, we do not analyze or generate any datasets. Our work proceeds within a theoretical and philosophical approach.
**Author Contribution**
All authors contributed to the study conception and design. Sections 1 and 4 were written with equal contribution. Sections 2 and 3 were conceived by Adomaitis and later edited by Grinbaum. Sections 5 and 6 were conceived by Grinbaum and later edited by Adomaitis. All authors read and approved the final manuscript.

# 1. Introduction

High-ranking authorities in Europe and the United States have recently expressed their concern regarding the societal implications of generative artificial intelligence (Coulter and Mukherjee 2023; White House 2023a). These concerns have quickly found their way into regulatory documents, e.g. the Executive Order signed by the United States President J. Biden (The White House 2023) or the recently amended and agreed-upon European legislative proposal for the AI Act (Council of Europe 2023; European Parliament 2023b). In academic circles, such worries have also been echoed by distinguished scientists, e.g. computer scientist Geoffrey Hinton compared the risks associated with generative AI to the regulation of biological and chemical weapons by emphasizing that, although not "foolproof", this regulation generally prevents their use (Heaven 2023). Perhaps the best examples of publicly declared concerns were, in early 2023, the call for a moratorium initiated by the Future of Life Institute and backed by figures such as Elon Musk; in late 2023, the "Bletchley Declaration" signed at the UK summit on AI (AI Safety Summit 2023). A temporary halt on the development of large language models (LLMs) "more powerful than GPT-4" (Future of Life Institute 2023) is not an isolated call, but part of a historical lineage of similar appeals in the face of emerging technologies, many falling under the concept of Dual Use Research of Concern (DURC). In this paper, we argue that DURC is relevant and applicable to the domain of generative AI, especially in relation to LLMs.

In Section 2, we introduce the concept of dual-use research, originally conceived in relation to chemical and biological weapons. Despite its roots, we illustrate that DURC fundamentally differs from this early interpretation of dual-use. Between 2005 and 2021, DURC has been employed in areas such as gain-of-function biological research and gene editing, maintaining its relevance as a utilitarian governance approach for potentially high-risk innovative biotechnological research. In Section 3, we propose the application of the DURC framework to generative AI, including Large Language Models (LLMs). We redefine the traditional DURC categories, typically used in biological research, to address dual-use concerns specific to generative AI and LLMs. In Section 4, we examine how DURC frameworks could play a role in establishing the shared responsibility of complex stakeholder networks that are often involved in the development and deployment of AI technologies. In Section 5, we explore the advantages and potential limitations of the DURC framework. Applying DURC to generative AI could heighten political awareness of its significant role as a societal transformation force. This symbolic value of DURC should not be underestimated, knowing that scientific research and ambitious technological innovation are not going to be put to a halt. We conclude in Section 6 by providing a series of specific recommendations for LLM research as an application of the DURC framework.

# 2. What is Dual Use Research of Concern?

In some regulatory frameworks (e.g. *Export Administration Regulations* 2013; *Regulation (EU) 2021/821* 2021), the notions of "dual-use" and "misuse" refer to the interplay between civil and military research. Dual use concerns initially came to the public eye at the time of the Manhattan Project, which led both to nuclear energy production and to the atomic bomb. Already at the time, the dual nature of the implications of their research plagued scientists and raised questions about what we now call "open science" (Schweber 2013). In biotechnology, the awareness of the dual use situation began with the recombinant DNA technology in the 1970s. In 1975, the Asilomar conference proposed a moratorium on genetic engineering (Berg et al. 1975). As we show below, this short-lived moratorium in biological research is similar to the current proposals for generative AI. In the later years, e.g. during the 2001 anthrax attacks in the United States, the dual-use debate went on to address concerns about potential use of research for bioterrorism (Atlas 2002). Once more, this is similar to the current debate on the dual use of AI (Urbina 2022).

There exists a more general sense in which technologies have been called "dual use", the widest scope being "technologies that could be used for either good or bad purposes" (Koplin 2023). This very broad definition seems to include numerous techniques and devices with multiple uses. For example, a kitchen knife is a necessary tool in everyday life but it also accounts for at least half of knife crimes (Hern et al. 2005). The mere possibility of using a technology for bad purposes is not enough for ethical analysis; one needs to consider the responsibility of the innovator and the imputability of the outcome. In the case of knife crime, it would be very strange to assign the responsibility to the manufacturer of the knife. Thus, the concept of dual use needs to be defined more precisely.

One possible criterion for classifying a technology as dual-use would be to require a demonstrable potential for large-scale harm via malevolent or negligent use of this technology. This introduces the aspects of scale and risk, as opposed to a mere possibility of one-shot effect. For instance, prior to World War I, the deployment of biological agents in warfare was rare and sporadic, typically involving immediate tactical measures such as well poisoning, launching diseased cadavers into besieged cities, or distributing infected blankets (Geissler and Moon 1999). The specific worry about the military use of toxins and pathogens emerged later, with research and application of typhoid and anthrax bacilli, cholera vibriones, dysentery bacteria, paratyphoid bacteria, and botulinus toxin, demonstrating the potential for large-scale harm. These biological and toxin weapons have led to the Biological and Toxin Weapons Convention (BTWC) and various verification projects (e.g. VEREX).

The considerations of scope and risk make the notion of dual-use more precise, and they point to specific frameworks of evaluating and addressing the risks of harm as well as responsibilities involved. In this paper, we choose to pursue an analogy between LLMs and an existing dual-use technology with a considerable history of DURC governance.

The current phase of the dual-use debate in biology was initiated in 2005 with the publication of a study reconstructing the 1918 Spanish influenza virus. In a note added in a late-stage revision of this publication, the authors inserted a new statement clarifying the purpose of their research. They stated that their work was driven by "historical curiosity" but also added a pragmatic goal: "The fundamental purpose of this work was to provide information critical to protect public health and to develop measures effective against future influenza pandemics" (Tumpey et al. 2005). The belated addition of such a utilitarian objective gave rise to a lively debate and a new wave of the DURC debate in biological research in the early 2000s.

A few years later, a similar scenario unfolded regarding gain-of-function (GOF) research on the H5N1 bird influenza virus. Searching for the genetic signature of pandemic transmission capabilities, scientists altered the H5N1 virus to render it airborne transmissible among mammals (Herfst et al. 2012; Imai et al. 2012). Once more, they justified their creation of an entirely new and potentially pandemic virus as essential for public health protection – a utilitarian argument that sparked intense debate in academic circles, with many scholars questioning its validity (Evans 2013; Lipsitch and Bloom 2012; Lipsitch and Galvani 2014). Critics voiced concerns about the creation of high and unprecedented risks in the pursuit of scientific knowledge, leading to suggestions and implementation of a research moratorium and publishing restrictions (Collins and Fauci 2012). Nonetheless, these publication restrictions only lasted a few months and were lifted after the original article had been revised. The federal funding freeze endured until 2017 and was lifted following the publication of the US regulatory framework on DURC (US HHS 2017).

This debate reveals a conflict between two paramount values in responsible research and innovation: the pursuit of knowledge and the safeguarding of public safety. At its core, the concept of DURC encapsulates a utilitarian dilemma, wherein a single scientific endeavor can simultaneously pose significant security threats and yield vital societal benefits. This dilemma is not exclusive to the fields of chemistry or biology; it can extend to other research areas, including artificial intelligence.

Nowadays, the most commonly accepted definition of DURC comes from the Fink Report: "Research that, on the basis of 'state of the art and knowledge,' could reasonably lead to knowledge, products or technologies that could be directly diverted and/or pose a threat to public health; agriculture, wildlife, flora, the environment and/or national security" (National Research Council 2004). This definition has been adopted in 2007 by the National Security Advisory Board for Biosecurity (NSABB), a body under the National Institutes of Health governing high risk in biological research (NSABB 2007). While the Fink Report and the NSABB guidelines are primarily oriented towards life sciences, there is no inherent aspect in the definition that would preclude its application to information technology and AI.

*Table 1. Distinction between the civil-military duality and dual use concerns in research.*

| Civil-Military Applications | Fink Report Dual Use |
| --- | --- |
| § 730.3 of US Export Administration Regulations (EAR): "A 'dual-use' item is one that has civil applications as well as terrorism and military or weapons of mass destruction (WMD)-related applications."<br><br>"The transfer from civil to military application involves a process in which social actors reinterpret the purpose of a technology from a peaceful to a hostile context" (Tucker 2012, p. 30). | "Research that, on the basis of 'state of the art and knowledge,' could reasonably lead to knowledge, products or technologies that could be directly diverted and / or pose a threat to public health; agriculture, wildlife, flora, the environment and / or national security" (National Research Council 2004). |

Both the "civil – military" dilemma and the "high risk – high benefit" utilitarian dilemma are called "dual use" in the literature. Here, we use the latter concept in line with recent scientific literature on DURC (Korn et al. 2019). There is no shortage of discussion of military applications of the life sciences, e.g. for making chemical weapons or bioterrorism. Digital technologies, including AI, have also been widely discussed in the context of civil vs. military application (Sanger 2023). There are concerns that AI-fueled cyberspace skirmishes could "escalate into conventional warfare" (Taddeo and Floridi 2018), as well as numerous scenarios explored for autonomous weapons and their role in warfare (Christie et al. 2023; Scharre 2018). While the military applications debate is broad and rich, there is little analysis of DURC aspects of generative AI research, and in particular of LLMs. However, in 2018, the Future of Humanity Institute have found that then-current AI technologies expand existing threats, introduces new threats, and changes the type of threats to the digital, physical, and political security of individuals and nations (Brundage et al. 2018). Five years later, OpenAI published AI governance commitments that heavily resound with concerns typical for dual-use research. Their first commitment is to safety "in areas including misuse, societal risks, and national security concerns, such as bio, cyber, and other safety areas", while they mention examples of ways in which systems "can lower barriers to entry for [bio, chemical, and radiological] weapons development, design, acquisition, or use" (OpenAI 2023a). Similarly, Anthropic made a dedicated study on the possibility of using LLMs for the design of biological weapons, highlighting growing biosecurity risks (Anthropic 2023). This discussion of risks can be seen as setting the stage for considering generative AI as dual use.

### 3. Applying DURC framework to generative AI and LLMs

In 2022 and 2023, the DURC approach is getting increasingly relevant to generative AI as LLMs are getting easier to build using standardized tools. This is similar to what happened in biology in the years following the publication of the Registry of Standard Biological Parts (Galdzicki et al. 2011). This modular approach promoted do-it-yourself construction of biological systems and gave easy access to the tools of synthetic biology. In generative AI, a group of researchers at Stanford recently published a method of obtaining a highly functioning model, known as Stanford Alpaca, which is "surprisingly small and easy/cheap to reproduce" (Taori et al. 2023). Another group of researchers managed to bring down the price of fine tuning LLMs even further (R. Zhang et al. 2023). Alphabet (formerly Google) has published a small LLM, called Gemini Nano (Pichai and Hassabis 2023), with the goal of running it on EDGE devices with little computational power.

There is a clear trend indicating that high-performance, finely-tuned LLMs are becoming increasingly available. Similar to the case of modular biological parts, LLMs can now be created inexpensively by a growing number of researchers. This increased accessibility amplifies the potential for misuse, necessitating the regulation of foundation models (Bommasani et al. 2022) due to their potential to spread misinformation, manipulate, influence, perpetrate scams, or generate toxic language. Specific social risks have been identified by LLM developers (Weidinger et al. 2021). While some of the identified risks can be emergent, i.e. unintended by the initial designers and arising from training, others can stem from poor design, e.g. insufficient mitigation against bias or poor elimination of toxic outputs. Further, output toxicity may be highly context-dependent, e.g. generating a phrase calling someone a dog might be an offense, or not, depending on the circumstances. This dependency on context is an intrinsic characteristic of several types of harmful language, including insults, medical advice, etc. In this sense, LLMs present risks that are highly contextual.

Some areas for potential misuse of LLMs carry a malevolent intention by design, e.g. the user's intention to make disinformation cheaper and more effective, or an intention to facilitate fraud, scams and more targeted manipulation, or to assist code generation for cyber-attacks, creating weapons, or illegitimate surveillance and censorship (Weidinger et al. 2021). These intentions can only come true in virtue of the LLM's capacity to empower such types of uses, which underscores the LLM potential for misuse. This potential is always present in the model itself, even if the user is unaware of it. Furthermore, an even more concerning aspect of LLMs lies with the 'unknown unknowns', i.e. potential misuses that are not foreseen by current foresight and risk evaluation benchmarks (Tamkin et al. 2021). This means that while recent efforts to mitigate the risks associated with LLMs have concentrated on specific types of harm, more harmful behaviour may emerge from future use of LLMs. Thus we believe that powerful generative AI research, including on LLMs, should be classified under DURC.

Before exploring the parallels between DURC in biological research and generative AI, we address an obvious difference between the two. Harmful biological agents usually exist in nature, whereas AI systems are works of human ingenuity. The former exist autonomously, independent of human influence, whereas the latter can be understood as agents only by projection. One might be tempted to infer that the risks stem from different sources: human intent versus naturally occurring life. However, the case of gain-of-function (GOF) research shows that this distinction is not necessarily valid: the variants of H5N1 virus manufactured for increasing scientific knowledge did not previously exist in nature. These artificially engineered forms of bird flu were intentionally created for research purposes, yet their creation has escalated pandemic risks. This example suggests that the boundary between life and technique does not imply an insurmountable difference between DURC categories in biology and in computer science.

Moreover, LLMs possess unique characteristics that distinguish them from other digital systems in terms of risk. First of all, they are considered a "generalist technology" (Vannuccini and Prytkova 2021), so instead of serving one purpose, they can generate an extensive range of outputs, which span across contexts of application. This generalization, combined with their autonomy, can create risks that are far less predictable and more pervasive compared to traditional digital systems. Secondly, LLMs have an exceptional ability to mimic human-like behavior (speech and appearances, in case of multimodal systems) and cognition. This poses unique risks for impersonation and individual exploitation. Lastly, LLMs accrue increasingly important roles in decision-making processes (logistics, healthcare, finance, etc.), which puts immense pressure on their reliability.

Research on artificial agents, be they biological or digital, could be subject to DURC. As part of the definition and regulation of DURC, NSABB identified a set of seven research categories as criteria for DURC (NSABB 2007). Although these categories have been devised for the life sciences, we adapt and rephrase them for the use with generative AI.

*Table 2. NSABB dual use categories applied to generative AI, including LLMs.*

| No. | NSABB Categories | Digital Dual Use Research of Concern |
|---|---|---|
| (1) | Enhances the harmful consequences of a biological agent or toxin. | While agency can be projected on AI systems by users, digital agents do not preexist in nature and do not possess ontological harmful properties like toxins. However, LLMs can be used for malicious activities, e.g. generating highly persuasive disinformation, creating deepfakes, or |
| | | enhancing cyberattacks (C and J 2023; Gregory 2022; Ropek 2023). Unlike biological agents, LLMs can both give rise to such activities and be used to improve the efficacy of human-designed activities with an explicit malicious intention. |
| (2) | Disrupts immunity or the effectiveness of an immunization without clinical and/or agricultural justification. | Rapid evolution of LLMs has drastically outpaced the development of countermeasures, such as content verification tools, watermarks, or fact-checking algorithms (Clark et al. 2021; Grinbaum and Adomaitis 2022b; Heikkilä 2022). It is increasingly challenging to distinguish between genuine and artificial content, rendering existing content moderation and recommendation systems ineffective (cf. "spin" attacks (Bagdasaryan and Shmatikov 2022)). LLMs can degrade the flow of language, including in important settings like computer code, legal texts, or medical statements, by inserting erroneous but difficult-to-detect flaws. This is not necessarily an intended purpose of LLM generation but an emergent property that is hard to control and thereby poses a significant threat. |
| (3) | Confers to a biological agent or toxin resistance to clinically and/or agriculturally useful prophylactic or therapeutic interventions against that agent or toxin or facilitates its ability to evade detection methodologies. | LLMs facilitate unpredictable and/or undetectable behaviors of digital systems. Transformer-based LLMs exhibit emergent behaviors without any obvious robust control mechanism (Wei et al. 2022). Models are being released without sufficient measures against model replication and potential inference attacks (Mireshghallah et al. 2022; Moradi and Samwald 2021). |
| (4) | Increases the stability of, transmissibility of, or ability to disseminate a biological agent or toxin. | LLMs can alter or modify computer code or human language to obfuscate malicious activity or intent. LLMs can be utilized to develop sophisticated obfuscation, cryptographic, or evasion techniques, making it difficult for security systems to identify or interpret attack vectors or actions of malicious agents (Oak 2022). The speed of generation exceeds human capacity to maintain conscious control of the proliferation of toxic or erroneous language. |
| (5) | Alters the host range or tropism of a biological agent or toxin. | The cost of deployment enhances the risks of dual use. in contrast with other mass-destruction weapons, "the materials and equipment required to create and propagate a biological attack using naturally occurring or genetically manipulated pathogens remain decidedly "low-tech," inexpensive, and widely available" (National Research Council 2007). The case of LLMs is even more severe since replicating a foundation model is accessible to individuals and the smallest of organizations (Taori et al. 2023; R. Zhang et al. 2023). LLMs already have the potential to revolutionize spear phishing and other types of attacks due to drastic reductions in cost and time (Hazell 2023). This availability drastically lowers the barriers to entry, and thus increases the range of actors that can engage in malicious uses. |

| (6) | Enhances the susceptibility of a host population. | LLMs are quickly becoming more accessible and widespread to all people speaking a language, as well as to programmers writing computer code. Professional groups and societies as a whole will increasingly become more reliant on LLMs. This dependence on AI-generated content and the erosion of trust in information sources can make abuses of AI systems more critical and consequential (Weidinger et al. 2021). |
|---|---|---|
| (7) | Generates a novel pathogenic agent or toxin or reconstitutes an eradicated or extinct biological agent. | LLMs can "invent" emerging capacities that lead to novel types of harms or toxic language. They can also reinforce known harms or attach vectors and apply them in novel applications. For example, LLMs can be used to automate cyberattacks, including phishing, mass-scale social engineering, and producing malicious code. By generating convincing content tailored to specific targets, LLMs make it easier for malicious actors to weaponize language (EUROPOL 2023). |

Based on the application of these criteria, LLMs and generative AI research can be considered as DURC. This categorization, however, does not imply that LLM research should be prohibited or that the benefits of the technology should not be exploited. Like in biology, where GOF research continues to this day, DURC raises awareness of risks and provides guidelines on how to encourage safety when applying LLM research.

Nuclear energy is another technological dual-use area that can be compared to LLMs (Koplin 2023). Apocalyptic projections apply in both cases: some estimate that nuclear risks could "set civilization back centuries" (Scouras 2019), while others claim that AI brings about existential risks (Bostrom and Yudkowsky 2014). However, the parallel between LLMs and nuclear science is more difficult to establish than the parallel with biological research. The biggest gap belongs with the problem of scarcity and accessibility of resources. A motivated and complex government effort is needed for nuclear science to be exploited with potential harm. Materials, such as uranium and plutonium, are rare, heavily regulated, and monitored by international bodies. Meanwhile, LLMs are either open source or can be developed with growing ease by private companies or even by individual researchers, as recent LLaMA-2 models show well (Touvron et al. 2023). While nuclear research requires political will at the level of countries, LLMs present a multi-stakeholder dilemma involving individuals, research labs, and business. This makes AI governance a more complex task requiring specific measures not seen in the nuclear sphere.

## 4. Dual-use and Shared Responsibility

The DURC framework is not a panacea but also not a pharmakon: it is a necessary pragmatic step in the governance of LLMs, similarly to what occurred in GOF research. Even if one agrees that LLM research is DURC, it does not by itself require setting a rigid regulatory framework; rather, it may suggest self-regulation measures like voluntary commitments (White House 2023b) or industry-wide self-governance bodies (OpenAI 2023b). In bioethics, there is a debate on whether a well-developed system of self-regulation could strike a better balance between respecting scientific openness and protecting society from harm. Some argue that it will work provided that scientists engage with the system in good faith (Resnik 2010); others oppose it, saying that it might be an overestimation of the scientists' competences in assessing security risks (Selgelid 2007). It remains to be seen if self-regulation can be efficient in the AI sector, not least because manufacturers are private companies which also need to promote their risky products.

One of the crucial functions of a DURC framework is to distribute shared responsibility along a complex chain of stakeholders in the case of malfunctioning or damage. In the case of LLMs, stakeholders include research and innovation actors such as individual researchers and engineering teams, industrial actors such as manufacturers or deployers of AI systems, ecosystem members such as open-source platforms and independent model evaluation groups, governance bodies including regulators and standardization agencies, and final users. To arrive at a working notion of responsibility, criteria need to be established for how it should be shared among the different actors in the value chain, e.g., the programmers who design a foundation model and those who design control layers, the trainers who select training data, the manufacturer of the AI system and that of possible plugins, an intermediary entity using the API supplied by the manufacturer, and the final user (Grinbaum et al. 2017). The HHS P3CO framework in biological or chemical research introduced a set of criteria and norms governing shared responsibility (US HHS 2017). Similarly, shared responsibility needs to be actively promoted in the AI sector. For example, the proposal for an "AI Liability Directive" from the European Commission establishes the responsibility of the manufacturer for each individual output of an LLM, even if the manufacturer could not possibly intend the system to produce such an outcome (European Commission 2022). A careful consideration of DURC should remove the burden of duality from the manufacturer and distribute it along the chain of actors involved in building up the dual or damaging aspect of a particular outcome, rather than the model itself. Similarly to GOF in biotechnology, researchers should not be liable for anything and everything that the system they have built might do in the future. Such indefinite precaution may lead to the "infantilization of technology" (Grinbaum and Groves 2013), which continues to treat technologies as direct 'organs' or 'extensions' of their creators, imputing them with unlimited responsibility. Instead, responsibility should be limited, much like the responsibility of parents for their children: LLMs should not be fixed in a position of eternal childspeak. This would set the stage for more equitable and accountable research from the societal perspective. With an accepted DURC framework, emergent issues of responsibility for the use, misuse, or outcomes of generative AI could be addressed systematically rather than ad hoc.

## 5. Benefits and limitations of the DURC framework

There are potential pitfalls to categorizing a research domain as DURC. One such risk is the erosion of open science benefits. Specifically, the implementation of DURC could obstruct or even preclude the online publication of AI models, disrupting the principles of open source and open data (LAION e.V 2023). If DURC constraints on the advancement of science become excessively restrictive, then less established or unverified researchers and research teams may struggle to pursue their work in a fully compliant way. Therefore, any proposed DURC framework for AI systems should strive to balance regulatory measures with the promotion of open source and open data in AI model development. While the application of DURC may inevitably curtail certain benefits, e.g. reproducibility, such limitations should be kept to a minimum.

The need to introduce the right amount of limitations and constraints leads to a known problem in DURC, namely excessive formalization and bureaucratization. When the NSABB published the final recommendations, they were adopted in policy documents, e.g. in the US Department of Health and Human Services "Framework for Guiding Funding Decisions about Proposed Research Involving Enhanced Potential Pandemic Pathogens" (Evans 2020; US HHS 2017), but overall they failed to make a lasting impact on science. One reason for this belongs with the use of language that can hardly, if at all, be implemented on the operational level: the guidelines suggested that scientists should consider scenarios that are "credible", "realistic", or "plausible", while also being "highly unlikely but still credible", based on fourteen different categories of possible damage. These requirements proved to be more suited for regulatory purposes or legal proceedings, than scientific work of biologists. As Evans puts it, "…mandating scientists [to] conduct research only [on] certain issues would be an unjustifiable burden on their freedom (in addition to any utilitarian assertions about the role of scientific freedom in promoting health outcomes)" (Evans 2020). Freedom of research, which is enshrined, e.g., in the German (Art. 5 Absatz 3 GG) or Austrian (Art. 17 StGG) constitutions, can be seen as a permanent counterweight to the utilitarian DURC arguments.

Moreover, as debates on GOF showed, the utilitarian analysis of risks and benefits is not clear-cut and can be manipulated. Some scientists think that "gain-of-function research can come in handy," while others admit that "their practical importance wasn't […] very extraordinary" (Dance 2021). In GOF research, the risks and benefits are subject to expert controversy and cannot be agreed upon consensually. The same applies to LLMs and generative AI models. An established analysis provides twenty-one identified types of harm (Weidinger et al. 2021), while the benefits of LLMs are difficult to quantify.

The fact that generative AI models are not designed with a specific purpose further complicates their benefit assessment within the DURC framework. As an example, Sam Altman, the CEO of OpenAI, has admitted to overlooking the 'problem-solving' criterion when building a business around a powerful technology that was not developed to provide a specific benefit or solve a particular problem (Mollman 2023). However, such limitations of the utilitarian approach are not unprecedented in the context of emerging technologies (Grinbaum and Groves 2013). It is important to consider the individual desire and the ambition of groundbreaking scientists. These qualities form an integral part of the virtue ethics analysis, which complements the utilitarian approach by considering the values of a scientist on an individual level.

Another type of limitation of the DURC framework is its focus on short- or, at most, medium-term horizon. Making realistic risk evaluations long into the future is not feasible due to uncertainty. LLMs, however, will have long-term effects on language (Grinbaum et al. 2021) which can hardly, if ever, be addressed via governance frameworks. For example, a language always carries, implicitly or explicitly, a particular set of cultural values that express civilizational choices and a particular mode of life. Over time, these values will influence the users of LLMs. Many such effects will remain invisible to the user, but their longer-term influence on language and culture as a whole will eventually emerge and therefore should not be ignored.

Moreover, DURC frameworks usually focus on state-funded research that can be misused to threaten public health or national security. DURC largely relies on policy levers directed towards government-funded research. These levers are, however, not in place for LLMs. Generative AI developers are typically funded independently, often through private corporations, and do not rely heavily on government support. The usual funding-related incentives and policy measures are not applicable to major LLM developers. Furthermore, the global and diffuse nature of LLM development means that regulatory efforts in one jurisdiction might not prevent misuse in another, not to mention the potential for clandestine development that evade law enforcement. On the whole, the implementation of DURC meets here new challenges that indicate a lack of "teeth" in any governance framework due to low entrance barriers and highly international character of generative AI.

Despite all these limitations, DURC has a positive, and sometimes necessary, role to play in research. One major role of the DURC framework is to facilitate the relationship between science and politics. The application of DURC to generative AI would reflect broad political awareness that LLMs are playing a major role in the life of society as a whole. LLMs are a powerful tool that can influence all aspects of life, from private to professional and political, and therefore create risks with far-reaching implications. The symbolic and political value of explicitly treating LLMs as DURC should not be overlooked. The DURC framework would set the stage and rules for reflecting upon, and anticipating, the influence of technology on society while addressing the inevitable conflicts that will arise.

6. **Specific DURC recommendations for generative AI and LLMs**

The benefits and limitations of DURC outlined in the previous section demonstrate the need to adapt this framework to generative AI and LLMs. Principles of risk governance should not be too general for DURC to be impactful and operational in generative AI. We suggest the following recommendations.

> 1a) Develop standardized benchmarks to evaluate foundation models and generative AI systems for intentional abuse
> 1b) Evaluate foundation models for unintentional harm via 'red-teaming' by independent human testers

Foundation models or general-purpose AI (GPAI) models, including openly accessible ones, have recently been included in the European AI Act (European Parliament 2023a) as requiring specific compliance measures. This makes them subject to regulation by public authorities. Other provisions of the AI Act only apply to AI systems understood as marketable products. Although the debates continue on which GPAI models should be regulated directly, involved parties agree on the importance of evaluating the AI systems that are accessible to the public (e.g., ChatGPT using GPT-3 or GPT-4). Evaluation is a cornerstone of DURC. For LLMs, it includes a variety of testing techniques (automatically computed benchmarks on standardized datasets, penetration testing, human 'red teaming', etc.). A DURC framework should begin with testing and evaluation requirements; the results of such testing should be published alongside the model.

2) Evaluate datasets and use high-quality data for training

Many emergent risks of LLMs stem from existing bias in training datasets, often replicating or amplifying harmful statistical associations in their training data (Caliskan et al. 2017; Qian et al. 2022). It is a problem that is particularly striking for historically marginalized groups, languages, and cultures (Field et al. 2021). Beyond bias, the epistemic quality of the training data is also an important factor, e.g. training on books vs. online forums leads to the outputs of vastly different quality. LLMs may also have the ability to "know what they don't know" (Osband et al. 2022). Depending on specific applications of the LLMs, they may be given additional requirements of both unfair bias mitigation, and adversarial quality control.

3) Enforce the human-machine distinction via watermarks

At a societal level, the use of nudging and deception can lend itself to political manipulation (Reisach 2021). LLMs can be used to create disinformation at scale (Xu 2020). A scalable production of fake content has the potential to create "filter bubbles" or "echo chambers", whereby media consumers rely only on unverified content (Colleoni et al. 2014). Moreover, chatbots can be designed to achieve optimal rankings in recommendation algorithms that supply the content to the end users, emphasizing specific political views. Risks arising from fraud or manipulation, which comprise a large spectrum of societal risks, imply that users should not mistake an output produced by a machine for an output created by a human author (Aaronson 2022; Grinbaum and Adomaitis 2022b; Kirchenbauer et al. 2023). The purpose of algorithmically designed watermarks is to maintain the possibility of distinguishing between machine and human authorship. Yet, the use of watermarking techniques for LLMs should remain unintrusive. The user's experience, e.g. in obtaining medical or legal advice, should not be perturbed by irrelevant disclaimers in the outputs. We recommend that watermarks be sufficiently hidden from the user but detectable only with a minor effort, as well as sufficiently robust to resist adversarial attempts to blur the origin of the text by editing. Even if watermark efficiency cannot be absolutely guaranteed (H. Zhang et al. 2023), the introduction of watermarks is a necessary regulatory step from the societal point of view.

4) Avoid overpolicing and sterilization of language

Any implemented controls must be proportional to the risks, while unnecessary limitations can distort or impoverish the generated language. The current iterations of LLMs encourage a stale and repetitive use of literary devices (Smith-Ruiu 2023) as well as an arbitrary avoidance of critical topics (Weidinger et al. 2021). Since humans imitate language abilities of their interlocutors, be they machines or other humans (Grinbaum 2023), this creates risks for the cognitive and cultural developments of the individual and of society as a whole. In general, output filtering is useful, for example generalist chatbots should not provide medical or legal advice. However, this has collateral effects, namely, excluding all types of "toxic" language results in a sterilized language. Human users may find that their own language gets less esthetically pleasing and more banal as a result of their interaction with chatbots.

5) Introduce a set of criteria and norms governing shared responsibility

LLMs and LLM-based chatbots are machines, yet they acquire qualities by projection (Grinbaum and Adomaitis 2022a). Their emergent "conduct" may lead to morally significant consequences. For example, chatbots can be perceived as lying, misleading, hurting, manipulating or insulting human beings (Davis 2016). Such effects usually induce respective projections of moral responsibility on machines. However, digital agents are not moral agents and cannot assume responsibility. A chatbot should never be perceived by the user as a responsible person, even by projection (Grinbaum 2019). To remove such projections, a comprehensive set of criteria and norms that clearly outline the shared responsibilities between AI developers, users, and other stakeholders should be created (Adomaitis et al. 2022; Dignum 2019). The DURC framework can help to phrase such criteria in a language understandable to legal professionals and non-experts in general. Shared responsibility should emphasize the collective dimension of design and deployment of AI systems.

In conclusion, the application of the Dual Use Research of Concern (DURC) framework to the field of generative AI and Large Language Models (LLMs) brings a new perspective on the rapidly expanding influence of these technologies. It serves as a call for reflective governance allowing researchers, policymakers, and the public to conscientiously address the broad-reaching implications of LLMs. The recommendations provided in this article are meant to offer tangible starting points for shaping the ethical trajectory of generative AI research, thereby ensuring that the balance between innovation and security is maintained. It is our hope that this perspective will stimulate further dialogue, leading to the emergence of robust strategies that uphold the integrity and potential of AI, while safeguarding societal interests.

# References

Aaronson, S. (2022, November 29). My AI Safety Lecture for UT Effective Altruism. *Shtetl-Optimized*. https://scottaaronson.blog/?p=6823. Accessed 11 January 2023

Adomaitis, L., Grinbaum, A., & Lenzi, D. (2022). *TechEthos D2.2: Identification and specification of potential ethical issues and impacts and analysis of ethical issues of digital extended reality, neurotechnologies, and climate engineering*. CEA Paris Saclay. https://hal-cea.archives-ouvertes.fr/cea-03710862. Accessed 25 October 2022

AI Safety Summit. (2023, November 1). The Bletchley Declaration by Countries Attending the AI Safety Summit, 1-2 November 2023. *GOV.UK*. https://www.gov.uk/government/publications/ai-safety-summit-2023-the-bletchley-declaration/the-bletchley-declaration-by-countries-attending-the-ai-safety-summit-1-2-november-2023. Accessed 23 December 2023

Anthropic. (2023, July 26). Frontier Threats Red Teaming for AI Safety. https://www.anthropic.com/index/frontier-threats-red-teaming-for-ai-safety. Accessed 31 July 2023

Atlas, R. M. (2002). National Security and the Biological Research Community. *Science*, *298*(5594), 753–754. https://doi.org/10.1126/science.1078329

Bagdasaryan, E., & Shmatikov, V. (2022). Spinning Language Models: Risks of Propaganda-As-A-Service and Countermeasures. In *2022 IEEE Symposium on Security and Privacy (SP)* (pp. 769–786). https://doi.org/10.1109/SP46214.2022.9833572

Berg, P., Baltimore, D., Brenner, S., Roblin, R. O., & Singer, M. F. (1975). Asilomar Conference on Recombinant DNA Molecules. *Science*, *188*(4192), 991–994. https://doi.org/10.1126/science.1056638

Bommasani, R., Hudson, D. A., Adeli, E., Altman, R., Arora, S., von Arx, S., et al. (2022, July 12). On the Opportunities and Risks of Foundation Models. arXiv. https://doi.org/10.48550/arXiv.2108.07258

Bostrom, N., & Yudkowsky, E. (2014). The ethics of artificial intelligence. *The Cambridge handbook of artificial intelligence*, *1*, 316–334.

Brundage, M., Avin, S., Clark, J., Toner, H., Eckersley, P., Garfinkel, B., et al. (2018). The Malicious Use of Artificial Intelligence: Forecasting, Prevention, and Mitigation. https://www.repository.cam.ac.uk/handle/1810/275332. Accessed 4 May 2023

C, D., & J, P. (2023, March 14). ChatGPT and large language models: what's the risk? *National Cyber Security Center*. https://www.ncsc.gov.uk/blog-post/chatgpt-and-large-language-models-whats-the-risk. Accessed 5 May 2023

Caliskan, A., Bryson, J. J., & Narayanan, A. (2017). Semantics derived automatically from language corpora contain human-like biases. *Science (New York, N.Y.)*, *356*(6334), 183–186. https://doi.org/10.1126/science.aal4230

Christie, E. H., Ertan, A., Adomaitis, L., & Klaus, M. (2023). Regulating lethal autonomous weapon systems: exploring the challenges of explainability and traceability. *AI and Ethics*, 1–17.

Clark, E., August, T., Serrano, S., Haduong, N., Gururangan, S., & Smith, N. A. (2021, July 7). All That's "Human" Is Not Gold: Evaluating Human Evaluation of Generated Text. arXiv. https://doi.org/10.48550/arXiv.2107.00061

Colleoni, E., Rozza, A., & Arvidsson, A. (2014). Echo chamber or public sphere? Predicting political orientation and measuring political homophily in Twitter using big data. *Journal of communication*, *64*(2), 317–332.

Coulter, M., & Mukherjee, S. (2023, April 17). EU lawmakers call for summit to control "very powerful" AI. *Reuters*. https://www.reuters.com/technology/eu-lawmakers-call-political-attention-powerful-ai-2023-04-17/. Accessed 10 May 2023

Council of Europe. (2023, December 9). Artificial intelligence act: Council and Parliament strike a deal on the first rules for AI in the world. *Press Release*. https://www.consilium.europa.eu/en/press/press-releases/2023/12/09/artificial-intelligence-act-council-and-parliament-strike-a-deal-on-the-first-worldwide-rules-for-ai/. Accessed 23 December 2023

Dance, A. (2021). The shifting sands of 'gain-of-function' research. *Nature*, *598*(7882), 554–557. https://doi.org/10.1038/d41586-021-02903-x

Davis, E. (2016). AI amusements: the tragic tale of Tay the chatbot. *AI Matters*, *2*(4), 20–24. https://doi.org/10.1145/3008665.3008674

Dignum, V. (2019). Taking Responsibility. In V. Dignum (Ed.), *Responsible Artificial Intelligence: How to Develop and Use AI in a Responsible Way* (pp. 47–69). Cham: Springer International Publishing. https://doi.org/10.1007/978-3-030-30371-6_4

European Commission (2022). Proposal for a DIRECTIVE OF THE EUROPEAN PARLIAMENT AND OF THE COUNCIL on adapting non-contractual civil liability rules to artificial intelligence (AI Liability Directive) (2022). https://eur-lex.europa.eu/legal-content/EN/TXT/?uri=CELEX%3A52022PC0496. Accessed 31 July 2023

European Parliament (2023a). Artificial Intelligence Act. Amendments adopted by the European Parliament on 14 June 2023. , Pub. L. No. P9_TA(2023)0236 (2023).

European Parliament (2023b). Artificial Intelligence Act: deal on comprehensive rules for trustworthy AI. Press release on 09 December 2023. https://www.europarl.europa.eu/news/en/press-room/20231206IPR15699/artificial-intelligence-act-deal-on-comprehensive-rules-for-trustworthy-ai

EUROPOL. (2023). *ChatGPT - the impact of Large Language Models on Law Enforcement*. Publications Office of the European Union. https://www.europol.europa.eu/publications-events/publications/chatgpt-impact-of-large-language-models-law-enforcement. Accessed 5 May 2023

Evans, N. G. (2013). "But Nature Started It": Examining Taubenberger and Morens' View on Influenza A Virus and Dual-Use Research of Concern. *mBio*, *4*(4), e00547-13. https://doi.org/10.1128/mBio.00547-13

Evans, N. G. (2020). Dual-Use and Infectious Disease Research. In M. Eccleston-Turner & I. Brassington (Eds.), *Infectious Diseases in the New Millennium: Legal and Ethical Challenges* (pp. 193–215). Cham: Springer International Publishing. https://doi.org/10.1007/978-3-030-39819-4_9

Export Administration Regulations (2013). https://eur-lex.europa.eu/legal-content/EN/TXT/?uri=celex%3A32021R0821

Field, A., Blodgett, S. L., Waseem, Z., & Tsvetkov, Y. (2021). A Survey of Race, Racism, and Anti-Racism in NLP. In *Proceedings of the 59th Annual Meeting of the Association for Computational Linguistics and the 11th International Joint Conference on Natural Language Processing (Volume 1: Long Papers)* (pp. 1905–1925). Presented at the ACL-IJCNLP 2021, Online: Association for Computational Linguistics. https://doi.org/10.18653/v1/2021.acl-long.149

Future of Life Institute. (2023, March 22). Pause Giant AI Experiments: An Open Letter. *Future of Life Institute*. https://futureoflife.org/open-letter/pause-giant-ai-experiments/. Accessed 3 April 2023

Galdzicki, M., Rodriguez, C., Chandran, D., Sauro, H. M., & Gennari, J. H. (2011). Standard Biological Parts Knowledgebase. *PLOS ONE*, *6*(2), e17005. https://doi.org/10.1371/journal.pone.0017005

Geissler, E., & Moon, J. E. van C. (1999). *Biological and Toxin Weapons: Research, Development, and Use from the Middle Ages to 1945*. Oxford University Press.

Gregory, S. (2022). Deepfakes, misinformation and disinformation and authenticity infrastructure responses: Impacts on frontline witnessing, distant witnessing, and civic journalism. *Journalism*, *23*(3), 708–729. https://doi.org/10.1177/14648849211060644

Grinbaum, A. (2019). *Les robots et le mal*. Paris: Desclée de Brouwer.

Grinbaum, A. (2023). *Parole de machines*. HUMENSCIENCES.

Grinbaum, A., & Adomaitis, L. (2022a). Moral Equivalence in the Metaverse. *NanoEthics*, *16*(3), 257–270. https://doi.org/10.1007/s11569-022-00426-x

Grinbaum, A., & Adomaitis, L. (2022b, September 7). The Ethical Need for Watermarks in Machine-Generated Language. arXiv. https://doi.org/10.48550/arXiv.2209.03118

Grinbaum, A., Chatila, R., Devillers, L., Ganascia, J.-G., Tessier, C., & Dauchet, M. (2017). Ethics in Robotics Research: CERNA Mission and Context. *IEEE Robotics & Automation Magazine*, *24*(3), 139–145. Presented at the IEEE Robotics & Automation Magazine. https://doi.org/10.1109/MRA.2016.2611586

Grinbaum, A., Devillers, L., Adda, G., Chatila, R., Martin, C., Zolynski, C., & Villata, S. (2021). *Agents conversationnels: Enjeux d'éthique* (Report). Comité national pilote d'éthique du numérique; CCNE.

Grinbaum, A., & Groves, C. (2013). What is "responsible" about responsible innovation? Understanding the ethical issues. *Responsible innovation: Managing the responsible emergence of science and innovation in society*, 119–142.

Hazell, J. (2023, May 12). Large Language Models Can Be Used To Effectively Scale Spear Phishing Campaigns. arXiv. https://doi.org/10.48550/arXiv.2305.06972

Heaven, W. D. (2023, May 2). Geoffrey Hinton tells us why he's now scared of the tech he helped build. *MIT Technology Review*. https://www.technologyreview.com/2023/05/02/1072528/geoffrey-hinton-google-why-scared-ai/. Accessed 3 May 2023

Heikkilä, M. (2022, December 19). How to spot AI-generated text. *MIT Technology Review*. https://www.technologyreview.com/2022/12/19/1065596/how-to-spot-ai-generated-text/. Accessed 5 May 2023

Herfst, S., Schrauwen, E. J. A., Linster, M., Chutinimitkul, S., de Wit, E., Munster, V. J., et al. (2012). Airborne Transmission of Influenza A/H5N1 Virus Between Ferrets. *Science*, *336*(6088), 1534–1541. https://doi.org/10.1126/science.1213362

Hern, E., Glazebrook, W., & Beckett, M. (2005). Reducing knife crime. *BMJ : British Medical Journal*, *330*(7502), 1221–1222.

Imai, M., Watanabe, T., Hatta, M., Das, S. C., Ozawa, M., Shinya, K., et al. (2012). Experimental adaptation of an influenza H5 HA confers respiratory droplet transmission to a reassortant H5 HA/H1N1 virus in ferrets. *Nature*, *486*(7403), 420–428. https://doi.org/10.1038/nature10831

Kirchenbauer, J., Geiping, J., Wen, Y., Katz, J., Miers, I., & Goldstein, T. (2023, January 27). A Watermark for Large Language Models. arXiv. https://doi.org/10.48550/arXiv.2301.10226

Koplin, J. J. (2023). Dual-use implications of AI text generation. *Ethics and Information Technology*, *25*(2), 32. https://doi.org/10.1007/s10676-023-09703-z

Korn, H., Pironneau, O., Fagot-Largeault, A., d'Artemare, B., Becard, N., Zini, S., et al. (2019). *RECHERCHES DUALES A RISQUE-Recommandations pour leur pris en compte dans les processus de conduite de recherche en biologie* (PhD Thesis). Académie des Sciences.

LAION e.V. (2023, May 4). An Open Letter to the European Parliament.

Lipsitch, M., & Bloom, B. R. (2012). Rethinking Biosafety in Research on Potential Pandemic Pathogens. *mBio*, *3*(5), e00360-12. https://doi.org/10.1128/mBio.00360-12

Lipsitch, M., & Galvani, A. P. (2014). Ethical Alternatives to Experiments with Novel Potential Pandemic Pathogens. *PLOS Medicine*, *11*(5), e1001646. https://doi.org/10.1371/journal.pmed.1001646

Mireshghallah, F., Goyal, K., Uniyal, A., Berg-Kirkpatrick, T., & Shokri, R. (2022, November 3). Quantifying Privacy Risks of Masked Language Models Using Membership Inference Attacks. arXiv. https://doi.org/10.48550/arXiv.2203.03929

Moradi, M., & Samwald, M. (2021, August 27). Evaluating the Robustness of Neural Language Models to Input Perturbations. arXiv. https://doi.org/10.48550/arXiv.2108.12237

National Research Council. (2004). *Biotechnology Research in an Age of Terrorism*. Washington, D.C.: National Academies Press. https://doi.org/10.17226/10827

National Research Council. (2007). Biosecurity and Dual-Use Research in the Life Sciences. In *Science and Security in a Post 9/11 World: A Report Based on Regional Discussions Between the Science and Security Communities*. National Academies Press (US). https://www.ncbi.nlm.nih.gov/books/NBK11496/. Accessed 11 May 2023

NSABB. (2007). Proposed framework for the oversight of dual use life sciences research: strategies for minimizing the potential misuse of research information. *A report of the National Science Advisory Board for Biosecurity (NSABB)*. National Science Advisory Board for Biosecurity, Office of Biotechnology ….

Oak, R. (2022). Poster – Towards Authorship Obfuscation with Language Models. In *Proceedings of the 2022 ACM SIGSAC Conference on Computer and Communications Security* (pp. 3435–3437). New York, NY, USA: Association for Computing Machinery. https://doi.org/10.1145/3548606.3563512

OpenAI. (2023a, July 21). Moving AI governance forward. https://openai.com/blog/moving-ai-governance-forward. Accessed 27 July 2023

OpenAI. (2023b, July 26). Frontier Model Forum. https://openai.com/blog/frontier-model-forum. Accessed 31 July 2023

Osband, I., Asghari, S. M., Van Roy, B., McAleese, N., Aslanides, J., & Irving, G. (2022, November 2). Fine-Tuning Language Models via Epistemic Neural Networks. arXiv. https://doi.org/10.48550/arXiv.2211.01568

Pichai, S., & Hassabis, D. (2023, December 6). Introducing Gemini: our largest and most capable AI model. *Google*. https://blog.google/technology/ai/google-gemini-ai/. Accessed 23 December 2023

Qian, R., Ross, C., Fernandes, J., Smith, E., Kiela, D., & Williams, A. (2022). Perturbation augmentation for fairer nlp. *arXiv preprint arXiv:2205.12586*.

Regulation (EU) 2021/821. , 206 OJ L (2021). http://data.europa.eu/eli/reg/2021/821/oj/eng. Accessed 10 May 2023

Reisach, U. (2021). The responsibility of social media in times of societal and political manipulation. *European Journal of Operational Research*, *291*(3), 906–917. https://doi.org/10.1016/j.ejor.2020.09.020

Resnik, D. B. (2010). Can Scientists Regulate the Publication of Dual Use Research? *Studies in Ethics, Law, and Technology*, *4*(1). https://doi.org/10.2202/1941-6008.1124

Ropek, L. (2023, January 20). ChatGPT Is Pretty Good at Writing Malware, It Turns Out. *Gizmodo*. https://gizmodo.com/chatgpt-ai-polymorphic-malware-computer-virus-cyber-1850012195. Accessed 5 May 2023

Sanger, D. E. (2023, May 5). The Next Fear on A.I.: Hollywood's Killer Robots Become the Military's Tools. *The New York Times*. https://www.nytimes.com/2023/05/05/us/politics/ai-military-war-nuclear-weapons-russia-china.html. Accessed 10 May 2023

Scharre, P. (2018). *Army of None: Autonomous Weapons and the Future of War* (First Edition, Stated, First Printing.). New York ; London: W. W. Norton & Company.

Schweber, S. S. (2013). In the Shadow of the Bomb: Oppenheimer, Bethe, and the Moral Responsibility of the Scientist. In *In the Shadow of the Bomb*. Princeton University Press. https://doi.org/10.1515/9781400849499

Scouras, J. (2019). Nuclear War as a Global Catastrophic Risk. *Journal of Benefit-Cost Analysis*, *10*(2), 274–295. https://doi.org/10.1017/bca.2019.16

Selgelid, M. J. (2007). A tale of two studies; ethics, bioterrorism, and the censorship of science. *The Hastings Center Report*, *37*(3), 35–43. https://doi.org/10.1353/hcr.2007.0046

Smith-Ruiu, J. (2023, April 5). My Dinners with GPT-4. *Justin Smith-Ruiu's Hinternet*. Substack newsletter. https://justinehsmith.substack.com/p/my-dinners-with-gpt-4. Accessed 10 May 2023

Taddeo, M., & Floridi, L. (2018). Regulate artificial intelligence to avert cyber arms race. *Nature*, *556*(7701), 296–298. https://doi.org/10.1038/d41586-018-04602-6

Tamkin, A., Brundage, M., Clark, J., & Ganguli, D. (2021, February 4). Understanding the Capabilities, Limitations, and Societal Impact of Large Language Models. arXiv. https://doi.org/10.48550/arXiv.2102.02503

Taori, R., Gulrajani, I., Zhang, T., Dubois, Y., Li, X., Guestrin, C., et al. (2023, April 3). Stanford Alpaca: An Instruction-following LLaMA Model. Python, Tatsu's shared repositories. https://github.com/tatsu-lab/stanford_alpaca. Accessed 3 April 2023

The White House. Executive Order on the Safe, Secure, and Trustworthy Development and Use of Artificial Intelligence (2023). https://www.whitehouse.gov/briefing-room/presidential-actions/2023/10/30/executive-order-on-the-safe-secure-and-trustworthy-development-and-use-of-artificial-intelligence/. Accessed 23 December 2023

Touvron, H., Martin, L., Stone, K., Albert, P., Almahairi, A., Babaei, Y., et al. (2023, July 19). Llama 2: Open Foundation and Fine-Tuned Chat Models. arXiv. https://doi.org/10.48550/arXiv.2307.09288

Tucker, J. B. (2012). *Innovation, Dual Use, and Security: Managing the Risks of Emerging Biological and Chemical Technologies*. MIT Press.

Tumpey, T. M., Basler, C. F., Aguilar, P. V., Zeng, H., Solórzano, A., Swayne, D. E., et al. (2005). Characterization of the Reconstructed 1918 Spanish Influenza Pandemic Virus. *Science*, *310*(5745), 77–80. https://doi.org/10.1126/science.1119392

US HHS. (2017). *Framework for Guiding Funding Decisions about Proposed Research Involving Enhanced Potential Pandemic Pathogens*. US Department of Health and Human Services.

Vannuccini, S., & Prytkova, E. (2021, April 7). Artificial Intelligence's New Clothes? From General Purpose Technology to Large Technical System. SSRN Scholarly Paper, Rochester, NY. https://doi.org/10.2139/ssrn.3860041

Wei, J., Tay, Y., Bommasani, R., Raffel, C., Zoph, B., Borgeaud, S., et al. (2022, October 26). Emergent Abilities of Large Language Models. arXiv. https://doi.org/10.48550/arXiv.2206.07682

Weidinger, L., Mellor, J., Rauh, M., Griffin, C., Uesato, J., Huang, P.-S., et al. (2021). Ethical and social risks of harm from Language Models. *arXiv preprint arXiv:2112.04359*.

White House. (2023a, May 4). Readout of White House Meeting with CEOs on Advancing Responsible Artificial Intelligence Innovation. *The White House*. https://www.whitehouse.gov/briefing-room/statements-releases/2023/05/04/readout-of-white-house-meeting-with-ceos-on-advancing-responsible-artificial-intelligence-innovation/. Accessed 10 May 2023

White House. (2023b, July 21). FACT SHEET: Biden-Harris Administration Secures Voluntary Commitments from Leading Artificial Intelligence Companies to Manage the Risks Posed by AI. *The White House*. https://www.whitehouse.gov/briefing-room/statements-releases/2023/07/21/fact-sheet-biden-harris-administration-secures-voluntary-commitments-from-leading-artificial-intelligence-companies-to-manage-the-risks-posed-by-ai/. Accessed 31 July 2023

Xu, A. Y. (2020, June 22). Creating Fake News with OpenAI's Language Models. *Medium*. https://towardsdatascience.com/creating-fake-news-with-openais-language-models-368e01a698a3. Accessed 30 August 2022

Zhang, H., Edelman, B. L., Francati, D., Venturi, D., Ateniese, G., & Barak, B. (2023, November 14). Watermarks in the Sand: Impossibility of Strong Watermarking for Generative Models. arXiv. https://doi.org/10.48550/arXiv.2311.04378

Zhang, R., Han, J., Zhou, A., Hu, X., Yan, S., Lu, P., et al. (2023, March 28). LLaMA-Adapter: Efficient Fine-tuning of Language Models with Zero-init Attention. arXiv. https://doi.org/10.48550/arXiv.2303.16199